\documentclass[prl,reprint,twocolumn,showpacs,notitlepage,groupedaddress]{revtex4-1}

\usepackage{graphicx}
\usepackage{amsfonts}
\usepackage{pdfsync}
\usepackage{amsmath}

\newcommand{\ket}[1]{|#1\rangle}

\newcommand{\hp}{|H\rangle}
\newcommand{\vp}{|V\rangle}
\newcommand{\spinup}{\mid \uparrow\rangle}
\newcommand{\spindown}{\mid \downarrow\rangle}

\begin{document}
\title{Reversal of Photon-Scattering Errors in Atomic Qubits}

\author{N.~Akerman}
\author{S.~Kotler}
\author{Y.~Glickman}
\author{R.~Ozeri}
\affiliation{Physics of Complex Systems, The Weizmann Institute of
Science, Rehovot 76100, Israel}

\begin{abstract}

Spontaneous photon scattering by an atomic qubit is a notable example of environment-induced error and is a fundamental limit to the fidelity of quantum operations. In the scattering process the qubit loses its distinctive and coherent character owing to its entanglement with the photon. Using a single trapped ion we show that by utilizing the information carried by the photon we are able to coherently reverse this process and correct for the scattering error. We further used quantum process tomography to characterize the photon-scattering error and its correction scheme and demonstrate a correction fidelity greater than 85\% whenever a photon was measured.

\end{abstract}
\pacs{42.50.Ct, 42.50.Dv}
\maketitle

%-------- Introduction ----------

%QEC of spontaneous emission
The process of spontaneous photon scattering by electronic superpositions in atoms is a prominent example of environment-induced decoherence. Error correction of spontaneous emission errors has been theoretically studied but mostly for qubit states that are separated by an optical frequency difference, and where photon emission is due to direct decay of the qubit excited state \cite{Plenio97,Lidar03}. Less attention has been devoted to the more common case where the qubit levels are both in the ground-state manifold, separated by the Zeeman or hyperfine interaction, and spontaneous emission errors occur by optical coupling of the qubit states to an excited level. Spontaneous scattering of photons by ground-state superpositions in atoms is an important source of error in many experimental systems. Examples include quantum information processing with trapped-ions \cite{Ozeri05, Ozeri07, Uys10} and coherent manipulations of ultracold gases in optical potentials \cite{Rosenfeld11}.

% Atom-photon entanglement
Spontaneous photon scattering entangles most spin superpositions with the scattered photon polarization \cite{blinov04, Weinfurter06,wilk07}. In cases where the photon is unmeasured, its degrees of freedom are traced out, leading to decoherence of the qubit. In contrast, whenever the photon is measured, this entanglement can be used as an information resource.
Such atom-photon entanglement was used to demonstrate the violation of Bell's inequality with an ion-photon pair \cite{Monroe04}, the heralded entanglement of two ions that were separated by one meter \cite{Monroe07}, as well as remote atomic state preparation \cite{Weinfurter07} and the heralded teleportation between two distant matter qubits \cite{Monroe09}. Here, we use atom-photon entanglement to correct for photon scattering errors.

% The level scheme in scattering process and general qubit dynamics due to photon scattering
A scheme of energy levels in $^{88}$Sr$^+$, which is relevant for the photon scattering process in our experiment, is shown in Fig.1(b). The qubit is encoded in the $\spinup = |m=+1/2\rangle$ and $\spindown = |m=-1/2\rangle$ electronic spin states of the $5\ ^2S_{1/2}$ ground level. Photon scattering is initiated by a weak, linearly polarized, laser beam that is on-resonance with the $5\ ^2S_{1/2}\rightarrow 5\ ^2P_{1/2}$ transition. Here, ion-photon interaction is governed by the electric-dipole interaction, $H_{int}=-\vec{d}\cdot\vec{E}$. Since $P_{1/2}$ is a spin 1/2 manifold as well, the effect of photon absorption or emission on the spin state can be described in terms of spin 1/2 operators. From the angular momentum part of the dipole interaction, we get that the coupling between the spin manifolds in the ground and excited orbitals is via $\vec{\sigma}\cdot\vec{E}$, where $\vec{\sigma}$ is a vector of the Pauli operators, indicating that the absorption and emission of a photon are associated with $\pi$ spin rotations around the photon electric field direction. Hence, given an initial spin direction, absorption of a photon transfers the electron to the $P_{1/2}$ orbital, while rotating its spin by $\pi$ around the absorbed photon polarization direction. The subsequent measurement of a linearly polarized emitted photon indicates that the electron has been  returned  to the $S_{1/2}$ orbital and that its spin has been rotated by $\pi$ around the detected photon polarization direction.

Formally, given an initial spin superposition, $\ket{\varphi}=\alpha\spinup+\beta\spindown$, a resonant excitation pulse, followed by spontaneous photon emission in the $\vec{k}$ direction, entangles the spin state of the ion with the polarization state of the emitted photon. The resulting ion-photon state can be written as,
\begin{align}
\label{Eq1}
\ket{\psi_{\vec{k}}}=\frac{1}{\sqrt{2}}(R_{\vec{\lambda}_1}&R_{\vec{\lambda}_L}\ket{\varphi}\otimes\ket{\vec{\lambda}_1} + R_{\vec{\lambda}_2}R_{\vec{\lambda}_L}\ket{\varphi} \otimes\ket{\vec{\lambda}_2}),\\
&R_{\vec{\lambda}}=\vec{\lambda}\cdot\vec{\sigma} = e^{i\frac{\pi}{2}\vec{\lambda}\cdot\vec{\sigma}}. \nonumber
\end{align}
Here $\vec{\lambda}_{L}$ is a unit vector in the excitation laser polarization direction, $\vec{\lambda}_{1,2}$ is a linear basis for the emitted photon polarization, and the lifetime in the excited state is neglected. The different photon frequencies that are associated with the different transitions due to their Zeeman shifts are ignored here since the time resolution of our photo-detectors introduces a frequency uncertainty much larger than this splitting and can be therefore considered as a quantum erasure for this degree of freedom \cite{Togan10}. In the absence of information on the emitted photon polarization, the ion spin state will decohere to a statistical mixture. However, with a measurement of the emitted photon polarization in a linear basis, the effect of photon scattering can be coherently reversed. In principle, the probability for successful correction can be very high and is only limited by the photon detection efficiency. Note that as long as the absorbed and detected photon polarizations are linear, photon detection carries information only on the way the spin evolved in the scattering process but no information on its pre  or post scattering states. On the other hand, if either the excitation laser polarization or the emitted photon detected polarization is not linear, the electronic spin undergoes a nonunitary time evolution, corresponding to a projective measurement, which cannot be reversed \cite{Glickman2012Emergence}. Formally if $\lambda$ is complex then the operator $R_{\lambda}$ in Eq.\ref{Eq1} is nonunitary and therefore irreversible.

%setup description
Our experimental setup is outlined in Fig.1(a). A single $^{88}$Sr$^+$ ion is trapped and Doppler cooled in a linear Paul trap. An externally applied magnetic field of $B$=1.25 G along the $\hat{z}$ direction defines a quantization axis and separates the ground state spin levels by $\omega_0/2\pi = 3.5$ MHz. The excitation laser beam propagates along a direction perpendicular to $\vec{B}$ (the $\hat{y}$ direction) and is polarized along $\hat{z}$ ($\pi$-polarized). Photons that are emitted along the $\hat{x}$ direction are collected by an objective lens with a numerical aperture of 0.31. The polarization of these photons is measured by two photomultiplier tubes (PMT's) on the different ports of a polarizing beam splitter (PBS). Quarter- and half-wave retardation plates, located in front of the PBS, allow one to measure the photon polarization in different bases. The total photon detection efficiency was found to be $\eta=2.5\times10^{-3}$. Each photon detection event produces two inputs to the control module which is based on a field-programmable gate array (FPGA) . One input indicates which PMT detects the photon and the second input records the phase of the local oscillator (LO), which is tuned to the qubit resonance frequency, at the time of detection. This phase is obtained using a time-to-amplitude converter (TAC) that measures the time interval between the PMT detection signal and the rising edge of the LO. Spin rotations are generated by means of radio frequency (rf) pulses. The ion spin state is detected using electron shelving followed by state selective fluorescence detection. See \cite{Akerman2011Quantum, Anna11} for more details.
\begin{figure}[h!]
\center
\includegraphics[width=9 cm]{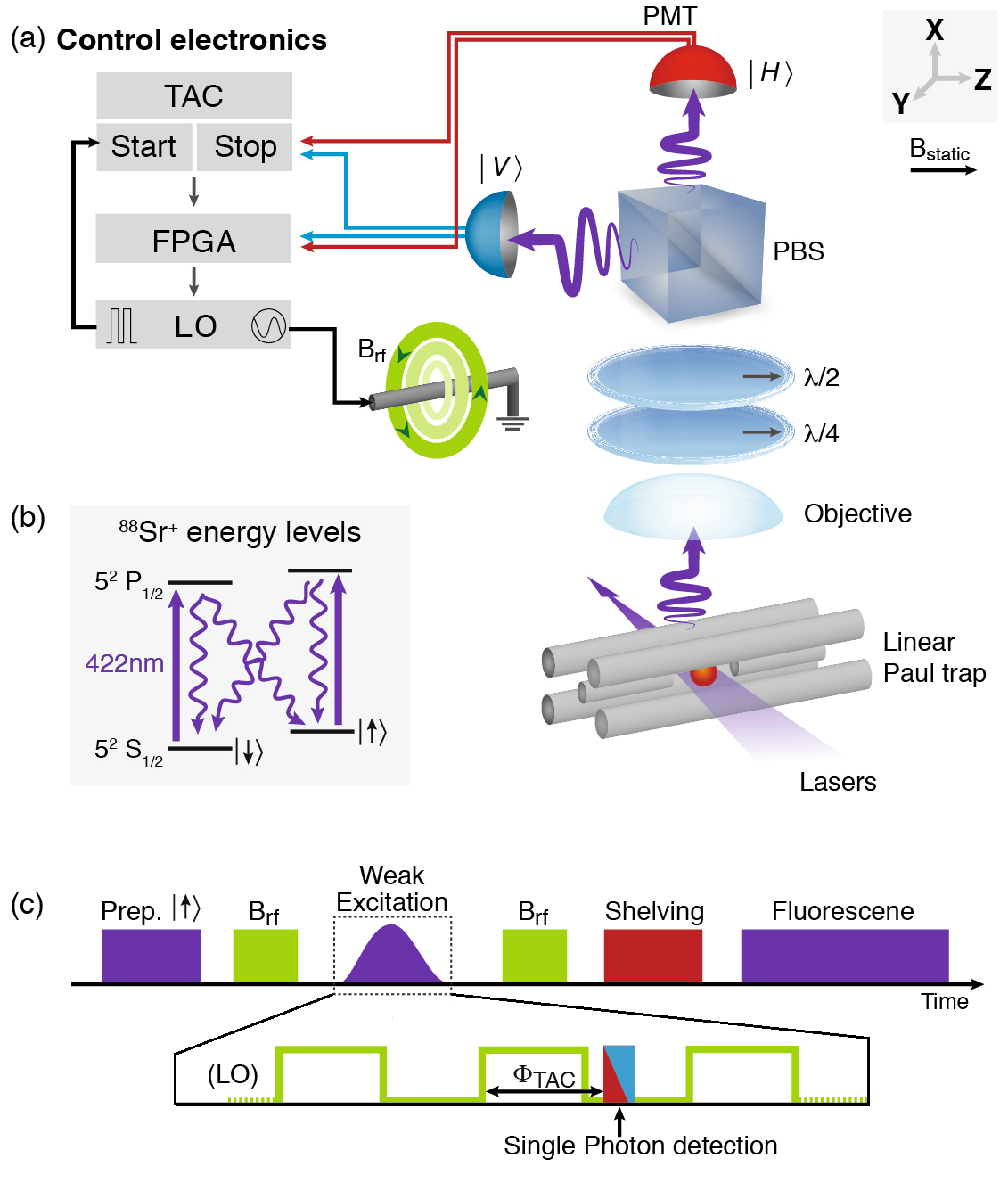}
\caption{ A schematic layout of our experiment. (a) Scheme of the apparatus. A single Strontium ion is trapped and laser cooled in a linear Paul trap. A weak $\pi$ polarized beam, resonant with the S$_{1/2}$ - P$_{1/2}$ transition, excites the ion which then spontaneously decays by emitting a photon. A photon collection apparatus consisting of a 0.31 N.A. objective lens, wave-plates, PBS and PMTs measures the polarization of photons that are emitted along the x direction. In addition, a TAC records the time interval between the rising edges of the LO and the PMT when a photon is detected. (b) A diagram of the  relevant energy levels in the $^{88}$Sr$^+$ ion. (c)  Experiment sequence. The spin is prepared in some state by optical pumping and rf spin rotation; then a photon is scattered by a weak resonant beam and the spin state is measured by another spin rotation, followed by shelving and state-selective fluorescence.}
\label{setupDrawing}
\end{figure}

%A general experimental sequence
The experimental sequence is illustrated in Fig.1(c). Preparation of the spin initial state is achieved by optical pumping followed by spin rotation. A weak 100-ns optical pulse at 422 nm subsequently excites the ion with a probability of 0.05-0.1. This sequence is repeated until a photon is measured. Depending on the measurement of a photon, the atomic spin state is measured in a chosen basis using spin rotations and detection. The complete scattering process dynamics is unveiled by performing quantum process tomography\cite{NielsenChuang}.
\begin{figure*}
\center
\includegraphics[width=16 cm]{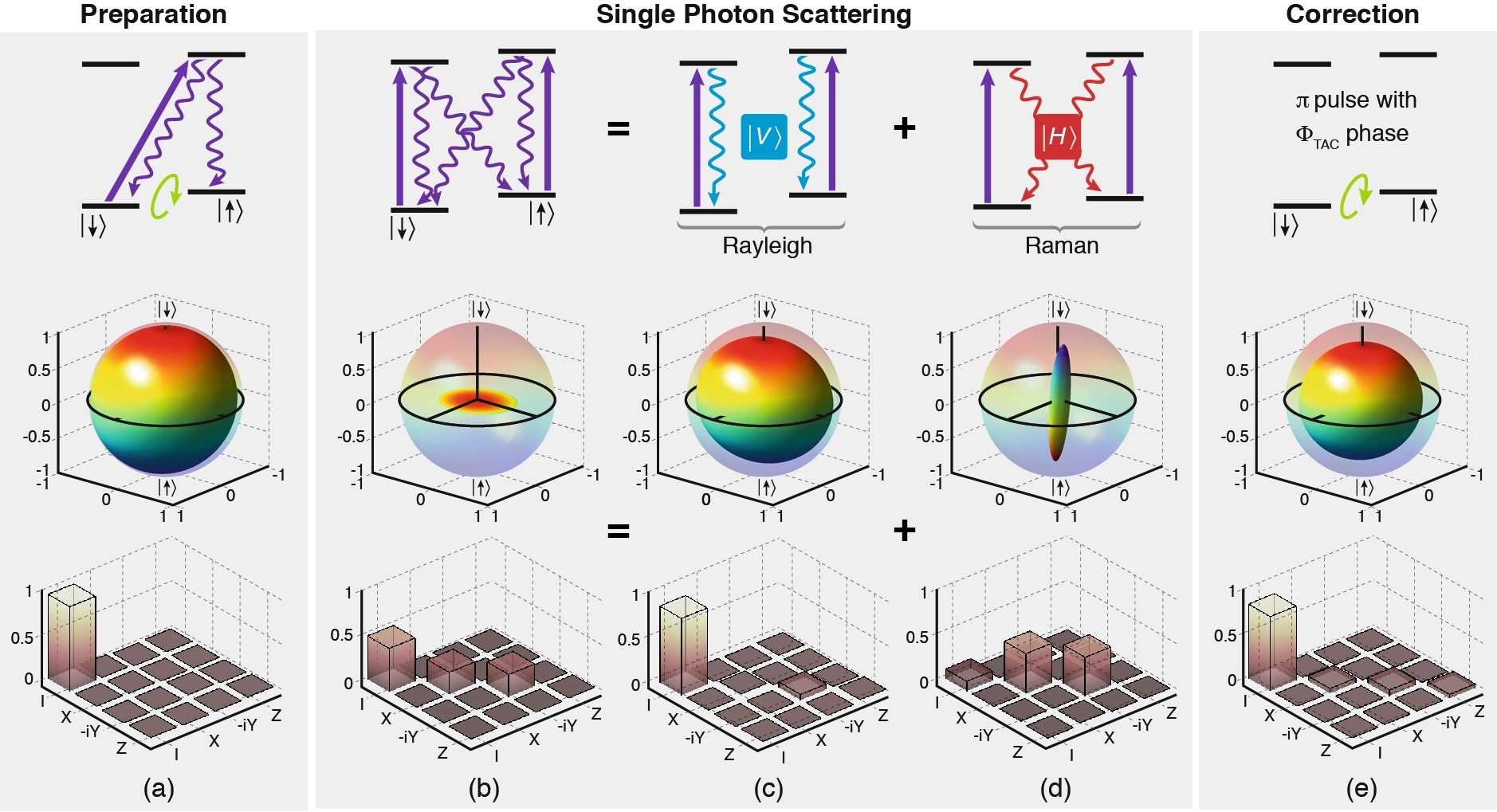}
\caption{Process tomography of the ion spin at the different stages of the experiment. The results are presented by the surface onto which all pure states are mapped and the real part of the reconstructed process matrix (the imaginary part is negligible and is not shown). (a) Without photon scattering. From the process matrix we infer qubit preparation and detection fidelity of 0.97. (b) After post-selection of single photon scattering in the $\hat{x}$ direction, photon scattering collapses the Bloch sphere of pure states onto a pancake-shaped spheroid in the equatorial plane. The process overlap with the ideal identity operation is reduced to 50\%. (c) and (d) Separating the measured result in (b), to $\vp$ (Rayleigh) and $\hp$ (Raman) photon detection events. Rayleigh scattering events leave the Bloch sphere almost intact, whereas Raman events swap between the sphere poles, while eliminating its width in the equatorial plane almost completely. Here, the loss of coherence is attributed to the random time of photon scattering events. (e) After applying a correction pulse that depends on $\hp$ photons detection and with a phase given by $\phi_{TAC}$, the resulting overlap with the identity operation increases to 83\%.}
\label{QEC_PT}
\end{figure*}

%Analyzing photon scattering errors
We began by investigating the decoherence of the ion spin state following photon scattering. Since, during the experiment, the spin precesses around the external magnetic field ($\hat{z}$), it is better described in a frame rotating synchronously with it. In this frame, Eq.1 is properly modified by replacing $R_{\vec{\lambda}}$ by a time-dependent rotation $\tilde{R}_{\vec{\lambda}}(t_s)=e^{i \sigma_z \omega_0 t_s} R_{\vec{\lambda}} e^{-i \sigma_z \omega_0 t_s}$, where $t_s$ is the photon scattering time.
To benchmark our ability to initialize and measure our spin qubit in different bases, we perform process tomography without photon scattering \cite{Anna11}. Figure 2(a) shows the process matrix in the absence of photon scattering and the surface onto which all pure spin states are mapped in the Bloch sphere picture. The dominant process is clearly the unity process (97\% overlap), whereas the resulting spheroid reasonably resembles the ideal Bloch sphere. Figure 2(b) shows the process matrix and the surface onto which all pure states are mapped, for events where a single photon was measured. The resulting process matrix is dominantly composed of an incoherent mixture of ~50\% unity process, and 25\% of $\pi$ rotations, each around the $\hat{x}$ and $\hat{y}$ directions. This process maps all pure states onto a pancakelike spheroid, with a radius of ~0.4 in the Bloch sphere equatorial plane, indicating a severe reduction in spin coherence. The spheroid width in the $\hat{z}$ direction is 0.05, consistent with quantum projection noise, and a complete mixing of spin populations.

% presenting the natural basis H and V ,Rayleigh and Raman
We then characterized the information carried by the scattered photon polarization and its implication for the spin state. This analysis is dependent on the choice of the basis of the polarization measurement. The first linear basis that we examined was $\ket{V}=\hat{z}$ and $\ket{H}=\hat{y}$. Photons polarized along $\ket{V}$ can only induce $\Delta m=0$ spin transitions and are therefore identified with elastic Rayleigh scattering. Photons that are polarized along $\ket{H}$ induce only $\Delta m=\pm1$ transitions and are therefore identified with inelastic Raman scattering (see the diagrams in the central upper part of Fig.2). Including absorption, the corresponding rotations are $R_{V} = \sigma_z\sigma_z = \hat{I}$ and  $R_{H} = \hat{\sigma}_y \hat{\sigma}_z = i\hat{\sigma_x}$. Here, the resulting ion-photon state in the rotating frame is
\begin{align}
\ket{\psi_{\hat{x}}}=&\frac{1}{\sqrt{2}}((\alpha\spinup + \beta\spindown)\otimes\vp+ \\
& i(\beta e^{-i\omega_0 t_s}\spinup + \alpha e^{i\omega_0t_s}\spindown)\otimes\hp),\nonumber
\label{Eq2}
\end{align}
Rayleigh scattering, therefore does not change the atomic spin, whereas Raman scattering  induces a $\pi$ rotation around an axis that rotates, with an angular frequency $\omega_0$, in the Bloch sphere equatorial plane. Figures 2(c) and 2(d) show the results of full process tomography for postselected Rayleigh and Raman photons respectively. Rayleigh scattering produces a process matrix with an 87\% overlap with the ideal unity operation, and leaves the Bloch sphere almost intact, whereas Raman scattering swaps between the Bloch sphere poles while shrinking its width in the equatorial plane to almost zero, as expected from complete loss of phase coherence. The loss of phase coherence here is because spin rotations, due to Raman scattering events, occur at random times, $t_s$.

% Rayleigh-Raman demonstrated in Ramsey
Correcting the scattering error  will not work unless we will account for the random times of the scattering event by measuring the phase $\phi_{TAC}=\omega_0 t_s$. To underscore the role of $\phi_{TAC}$ on the spin, we perform a Ramsey experiment in which a photon is scattered between two $\pi/2$ rotations around the $\hat{x}$ axis, implemented with in-phase rf pulses and the results are sorted by their recorded $\phi_{TAC}$. As shown in Fig.3(a), Rayleigh scattering is independent of $\phi_{TAC}$, whereas in Raman scattering events, a $\pi$ rotation around the rotating axis produces a high-contrast double fringe.% 125 word

In the next step, we implement a correction scheme that consists of an rf $\pi$ pulse that is dependent on detecting an $\hp$ photon and with a phase that is determined by the TAC output. This pulse rotates the spin back to its initial state and reverses the effect of photon scattering. We characterized our correction scheme using quantum process tomography. Figure 2(e) presents the results of process tomography following correction. The reconstructed process matrix shows an 83\% overlap with the identity process. The resulting sphere onto which all pure spin states are mapped indicates that the collapse of the Bloch sphere due to photon scattering is largely suppressed. Few technical imperfections contribute to the infidelity of our correction scheme. These include multiple photon scattering ($3\%-7\%$), dark counts of the PMTs ($\sim3\%$), spin state preparation and detection errors ($\sim3\%$), photon polarization measurement inaccuracy due to the finite solid angle ($\sim1\%$), inaccuracies of the polarization analysis setup($\sim1\%$) and uncertainty in the duration that the spin spends in the excited state ($\sim1\%$).
\begin{figure}
\center
\includegraphics[width=8.9 cm]{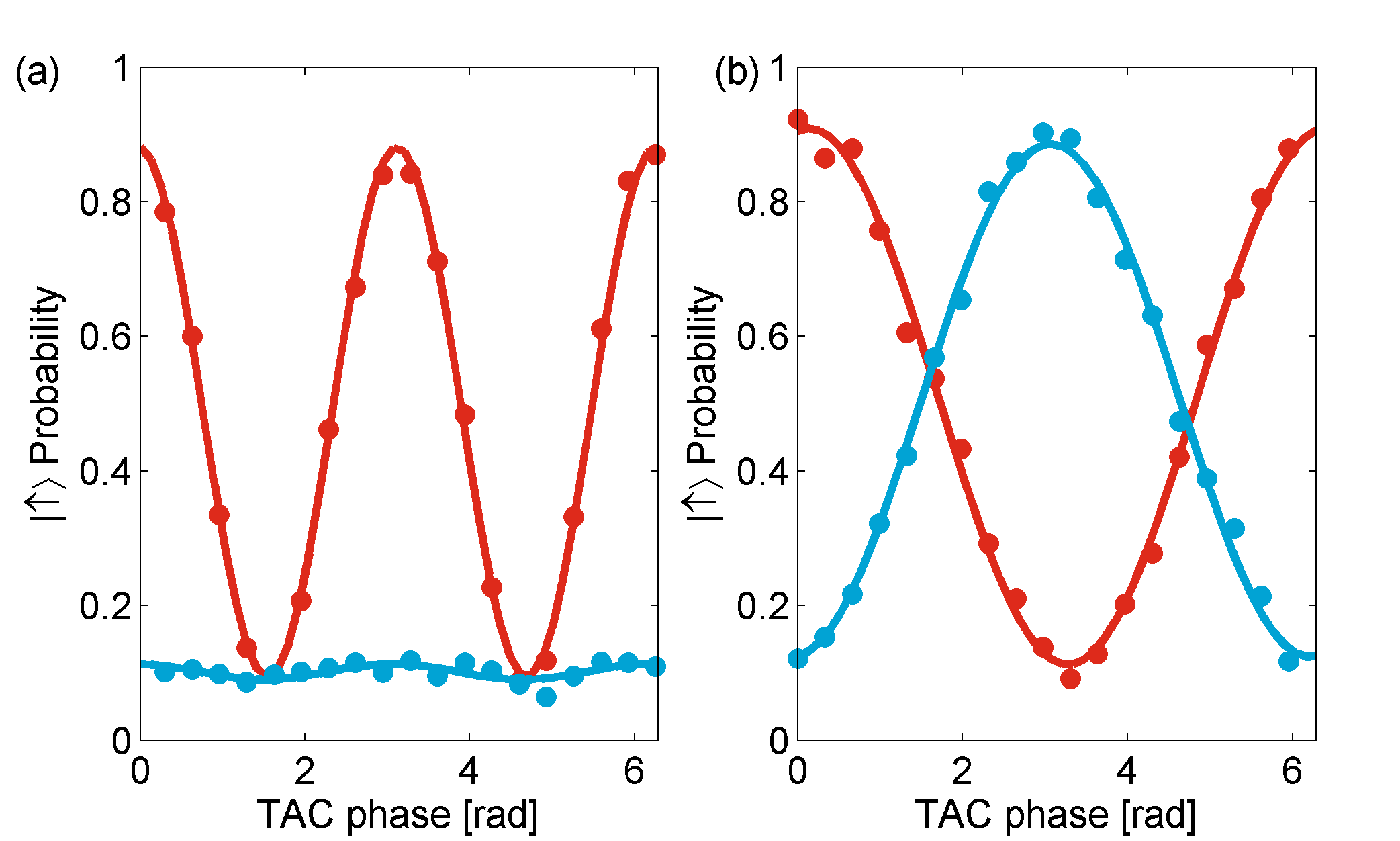}
\caption{A Ramsey experiment on the ion spin with single-photon scattering. (a) Measuring the  photon polarization in the basis $\ket{H},\ket{V}$, Probability to detect the spin in the $\spinup$ state vs. the TAC phase, for Rayleigh (blue) and Raman (red) scattering events separately. The probability of measuring $\spinup$ is independent of $\phi_{TAC}$ for Rayleigh scattering, whereas for Raman scattering, the $\pi$ rotation around an axis that rotates in the Bloch sphere equatorial plane, produces a double fringe. (b) Measuring the photon polarization in the basis $\ket{H'}=(\ket{H}+\ket{V})/\sqrt{2}$ and $\ket{V'}=(\ket{H}-\ket{V})/\sqrt{2}$. Here the $\pm\pi/2$ rotation due to the photon scattering in the two polarization $\ket{V'}/\ket{H'}$ (blue/red) substitute the first Ramsey pulse. The small phase shift (0.5 rad on top the $\pi$) is due to uncompensated birefringence in our polarization analysis setup.}
\label{Ramset45}
\end{figure}

% linear 45 demonstration in Semi-Ramsey
Using a different linear polarization measurement basis for the photon causes the ion to undergo different rotations. To study this effect, we chose a second polarization basis that is rotated by $45^{\circ}$ with respect to $\ket{H}$ and $\ket{V}$: $\ket{H'}=(\ket{H}+\ket{V})/\sqrt{2}$ and $\ket{V'}=(\ket{H}-\ket{V})/\sqrt{2}$. Here, following the detection of a ($\ket{V'}$ ) $\ket{H'}$ photon, the ion spin is rotated by an angle of $\pi/2$ (counter) clockwise around the $\hat{x}$ axis, which, in the rotating frame, rotates in the Bloch sphere equatorial plane. The evidence for the $\pi/2$ rotation is presented in Fig.3(b). Here again, a Ramsey experiment was performed but now the first Ramsey $\pi/2$ pulse, with a phase given by $\phi_{TAC}$, is induced by a photon scattering event. In the experiment the ion spin was initialized to $\spinup$, a photon was scattered, and then a $\pi/2$ rotation around the $\hat{x}$ direction was applied. As expected,  high contrast and opposite phase Ramsey fringes are observed for $\ket{V'}$ (blue) and $\ket{H'}$ (red) photon scattering events. From the fringe contrast in Fig.3(a) and Fig.3(b) we found that the entanglement fidelity, i.e., the overlap of the density operator we observed in this experiment and the ideal state in Eq.2, was $F>84 \%$ \cite{blinov04}. Next, we performed quantum error correction using photon measurements in the $\ket{H'}$, $\ket{V'}$ polarization basis. Here, the correction pulses were $\pi/2$ or $-\pi/2$ rotations with a phase given by $\phi_{TAC}$. Process tomography of our correction protocol yielded an error correction fidelity of 85\%. %350 words

%----------- Summary ------------
In conclusion, we have experimentally studied the dynamics of ground-state spin qubit due to the photon scattering process. We have shown that photon scattering, when restricted to linear polarization, generates $\pi$ rotations around the excitation laser and the emitted photon polarizations. We have further demonstrated that measuring the scattered photon polarization in a linear basis, allows for the reversal of photon scattering error. While here photon scattering is artificially introduced, it simulates real experimental situations where photons are scattered from linearly polarized laser beams such as spontaneous scattering from an off-resonance optical dipole trap beam. As in other heralded schemes in quantum information science, a practical use of our demonstrated technique demands a substantial improvement of the photon detection efficiency.

The error model that we have characterized here is comprised  of all uncorrelated single qubit rotations; thus the full 5-qubit quantum error correction code (QECC) \cite{laflamme1996perfect} is needed in order to protect a ground state qubit against spontaneous scattering errors. This is in contrast to the eight-qubits QECC needed to protect against the spontaneous decay of the qubit excited state \cite{Plenio97}. As we have shown here, whenever, information about the scattering process is obtained, the resources needed for QECC can be relaxed. If the qubit from which scattering occurs is known, then in the erasure channel, four qubits suffice to correct a general error \cite{Grassl97}. If the direction and time of photon scattering are known as well, a correction of a $\pi$ rotation around a known direction is required. Here, a two-qubit code and parity check will be sufficient. As we have shown here, with a measurement of the scattered photon polarization no ancillary qubits are needed and the error can still be reversed.

This research was supported by the Israeli Science Foundation, the Minerva foundation, the German-Israeli Foundation for scientific research, the Crown Photonics Center, the Wolfson Family Charitable Trust, Yeda-Sela Center for Basic Research and David Dickstein, France.

\end{document}